\documentclass[12pt,a4paper]{article}
\usepackage{graphicx}
\usepackage{cite}

\begin{document}

\title{Gas dynamic simulation of the star-planet interaction \\ using a binary star model}

\author{
       D.\,E.\,Ionov,
       D.\,V.\,Bisikalo,
       P.\,V.\,Kaygorodov,
       V.\,I.\,Shematovich\\[20pt]
\small{Institute of Astronomy of the Russian Acad. of Sci.}\\[1pt]
\small{Pyatnitskaya 48, Moscow, Russia}
} 

\date{2012}

\maketitle
\begin{abstract}
We have performed numerical simulations of the interaction between
``hot Jupiter'' planet and gas of the stellar wind using numerical
code developed for investigations of binary stars. With this code
we have modeled the structure of the gaseous flow in the system
HD~209458. The results have been used to explain observations of
this system performed with the COS instrument on-board the  HST.

\end{abstract}

The observation of a "hot Jupiter" planet HD~209458b were carried
out using the COS spectrograph mounted aboard HST \cite{Linsky}. The results showed that the investigated spectral
absorption lines (of CII, Si III) obtained as the difference of
the stellar spectra in the transition and out of it have a
non-trivial double-peaked shape. The distance between the peaks is
about 20~km/s and for the carbon line it is clearly seen that the
peaks are asymmetric.

To reveal physical processes
that can lead to formation of such spectral lines we performed gas
dynamic simulations of the interaction between a planet and
stellar wind. Since the planet is pretty close to the host star
(10.1 $R_{sun}$), the orbital velocity of the planet is so high
(V=143 km/s) that even in a case of a hot stellar wind with
temperatures around $10^5$ K the motion of the planet is
supersonic. It is known that if a gravitating body or a body with
an atmosphere moves with a supersonic velocity a bow-shock must
occur. The matter of the stellar wind mixes with the matter of the
athmosphere and forms two streams moving in different directions
from the head-on collision point. This motion can lead to
occurrence of two peaks in observed spectral lines.

The model we have used for our numerical experiments is similar to
that described in~\cite{Bisik03AZ}. The flow structure in
this model is described by a system of 3D equations of
gravitational gas dynamics including non-adiabatic processes of
the radiative heating and cooling. To obtain the numerical
solution of this system we have used a Roe-Osher method~\cite{Masstr,Roe,Chakravarthy, Bisik03MM} adapted to perform simulations using multiprocessor
computers. We carried out our calculations in a cylindrical
coordinate system with the origin in the center of mass of the
planet. The size of the computational domain has been limited by 5
planet radii. The stellar wind was simulated through setting a
constant inflow on the outer boundary of the computational domain.
The matter of the stellar wind is considered as a neutral
monatomic gas whose parameters are typical for the Solar wind,
i.e.: $\rho=1,4 \cdot 10^4\,cm^{-3}, T=7,3 \cdot 10^5\, K, V=100\, km/s$~\cite{Withbroe}. The planet atmosphere is set
isothermal with $T_{pl} = 5000\, K$. Its number density has been
determined using the barometric formula.  The number density at
the radius $R_{pl}=1.4 R_{Jup}$ has been set as $2 \cdot 10^{10}\,
cm^{-3}$.

\begin{figure}[t]
\begin{center}
\begin{tabular}{cp{0.5cm}c}
 \includegraphics[width=2.7in]{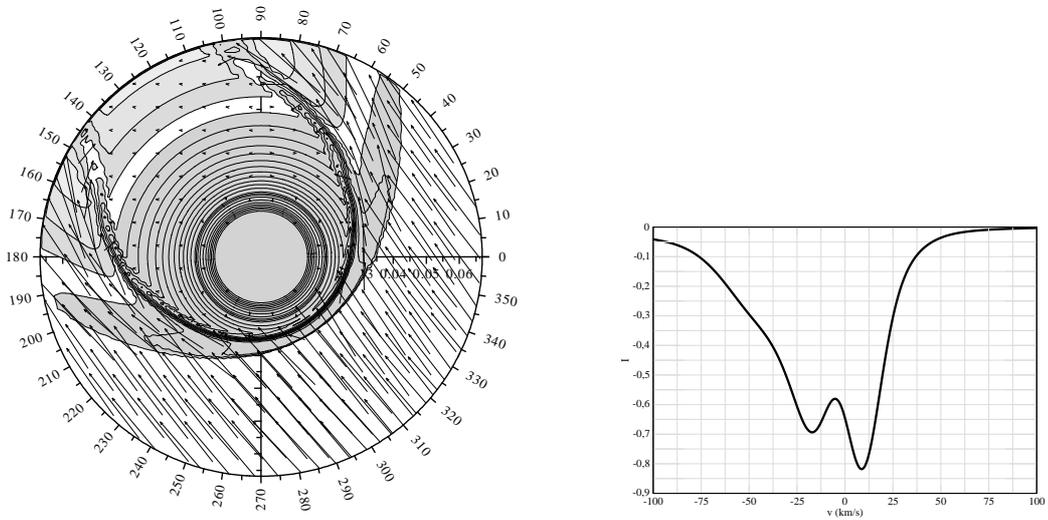}&&\includegraphics[width=2.2in]{fig2a.eps}
\end{tabular}
\caption{Density distribution and velocity vectors in the
equatorial plane (left panel) and synthetic spectral line
profile(right panel).}
   \label{fig1}
\end{center}
\end{figure}

A shape of the line is determined by projections of the velocities
of matter behind the bow shock wave onto the line of sight. In
Fig.\,\ref{fig1} (right panel) we show synthetic absorption line
profile calculated using our gas dynamic results. It is seen that
the line has two distinguishable clearly asymmetric peaks. The
synthetic line profile has the same features as the observed one.
Parameters of the line (peak position, their relative intensity
and even their number) strongly depend on accepted parameters of
the modeled stellar wind and athmosphere, and can vary in a wide
range.

The presented model allows us to conclude that when
analyzing observational properties of the atmosphere of a planet
one must take into account gas dynamical processes caused by the
interaction of the atmosphere and stellar wind. It is important to
note that the considered model allows one to explain the existence
of absorption lines of ions having high ionization potentials.

\bigskip

{\bf Acknowledgements.}
 This work was supported by the Basic Research Program of the
Presidium of the Russian Academy of Sciences, Russian Foundation
for Basic Research (projects 09-02-00064, 09-02-00993,
11-02-00076, 11-02-01248), Federal Targeted Program "Science and
Science Education for Innovation in Russia 2009-2013".

\end{document}